\documentclass[smus]{snow2e}
\usepackage{epsf}
\newcommand{\eg}{{\it e.g.}}
\newcommand{\etal}{{\it et al.}}
\begin{document}

\title{Future Directions for QCD\thanks{Work supported by 
the Department of Energy, contract number DE--AC03--76SF00515.}}

\author{James D. Bjorken\\ {\it Stanford Linear Accelerator Center,
                                Stanford, California 94309}}

\maketitle

\thispagestyle{empty}\pagestyle{empty}

\begin{abstract} 

New directions for exploring QCD at future high-energy colliders are
sketched.  These include jets within jets, BFKL dynamics, soft and
hard diffraction, searches for disoriented chiral condensate, and doing
a better job on minimum bias physics.  The new experimental
opportunities include electron-ion collisions at HERA, a new collider
detector at the C0 region of the TeVatron, and the FELIX initiative at
the LHC.

\end{abstract}

\section{Introduction}
\label{sec:intro}

This talk is not meant to be a comprehensive overview of QCD. The
emphasis here is simply on those aspects of QCD theory and
phenomenology most relevant to the Snowmass mission, namely, (1)
new-facility opportunities, (2) new, relatively unexplored, directions
in QCD theory and/or experiment, and (3) the difficult areas of QCD
which need to be data driven, but where the data is insufficient.

Before entering into these somewhat specific and perhaps idiosyncratic
topics, it must be put on the record that a large core region of theory
and phenomenology is in quite good shape, and quite mature. The QCD
Lagrangian has been ``tested" so incisively that few if any theorists
now challenge the correctness of the QCD Lagrangian (this includes
yours truly). Yes, $\alpha_s$ runs. The parton structure of hadrons and
much of hard-collision phenomenology are well understood. Extrapolation
to the higher energies and new facilities can be done with confidence,
at least at the level needed for design purposes.

However, there are many fundamental issues in QCD which are not
in good shape. Quite a few involve the low-energy nonperturbative
sector, \eg\  the question of confinement, and do not meet the criteria
for inclusion in this talk as outlined in the first paragraph. My views
of some of these are covered in another talk, given at the SLAC Summer
Institute this year\cite{ref:A}.

The QCD physics issues which will be discussed in the next section
include

\begin{Itemize}
\item
Fractal final-state phase space
\item
Black quarks
\item
Soft and hard diffraction
\item
The chiral phase: disoriented chiral condensate
\item
Underlying-event and minimum-bias physics. 
\end{Itemize}

In Section~\ref{sec:3} we discuss these topics in the context of physics 
opportunities at new machines and/or new facilities at old machines.

\section{Physics}
\label{sec:2}

\subsection{Fractal final-state phase space}
\label{ssec:2a}

The final-state phase space in the high-energy, high-$p_t$ limit of
strong interactions is fractal\cite{ref:B}.  By this we mean that the
QCD branching structure of parton cascades leads to jets within jets
within jets$\ldots$\ Each jet extends the phase-space region into which
hadrons are produced. Because of the self-similar nature of the parton
cascade, this leads to an anomalous dimension of the phase space.

To get some more concrete appreciation for what the above words are
supposed to mean, consider a typical Fermilab Tevatron multijet final
state, with the jets well scattered in the lego plot. In the detector
most of these jets will be at small angles. But suppose that the system
were produced at large angles instead. Then in the lego plot, with the
usual Snowmass-accord definition of jets, the multijet configuration
would most likely be able to be described as a mere two-jet final
state. The original information of the multijet textures of the
right-moving or left-moving systems would be compressed into single
circles-of-radius-0.7. Clearly one should not be content with such a
description. It is however easy to retrieve the original information
without an overall coordinate rotation\cite{ref:C}. One simply
introduces polar coordinates inside each Snowmass circle, trades the
new polar angle in for an appropriately defined rapidity variable and
replots the contents of the interior of the circle into a new lego
plot. Note that the area of the new lego plot will be $2\pi\log p_t$.
If jets are found in the new lego plot the process is iterated until
none remain. Then the mean density of produced hadrons in this extended
lego plot can be expected to be rather uniform, and therefore the total
multiplicity can be expected to be proportional to the total lego area,
which clearly has fractality built in.
 
To see this fractality clearly is a big experimental challenge. For the
leading-jet systems prevalent in hadron-hadron collisions, the
calorimeter resolution in the forward direction is made very good; the
pixel area (in real space) near the beam axis is made small, so that
\eg\ the number of pixels per unit azimuthal angle does not depend upon
distance from the beam. For the rotated jet systems, one needs to
accomplish this in all directions.

Thus the frontier becomes very small pixel size, when expressed in lego
variables, everywhere in the detector. In practice a $\delta\eta \times
\delta \phi$ of $0.003 \times 0.003$ might be attainable for resolving
charged tracks and $\gamma$'s, and this would give good resolution in
the first phase-space extension out to $\eta^\prime$ of about 4. This
would require of order $10^6$ pixels for a typical $4\pi$ detector.

\begin{figure}[htb]
\leavevmode
\begin{center}
{\epsfxsize=3in\epsfysize=1.75in\epsfbox{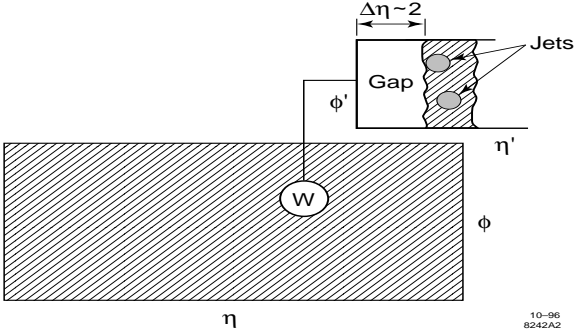}}
\end{center}
\caption{Extended phase-space for high-$p_T$ $W$ production.}
\label{fig:d}
\end{figure}

Might such a capability be useful beyond QCD? It seems to me that the
answer is affirmative. For example, one might search for rapidity gaps
in transverse, extended phase space. For example consider a $W$
produced at $p_t = 800\ GeV$ decaying symmetrically into two jets. The
separation of the jet cores is 0.2 radians; in the extended phase-space
the jet cores will appear at $\eta^\prime = 2$ (Fig.~\ref{fig:d}).
And there will be a
rapidity gap in the extended-phase-space lego plot. This feature may be
useful for production of color-singlet objects which are even more
interesting than a $W$.

\subsection{Black Quarks}
\label{ssec:2b}

Quark-parton interactions at a large, fixed momentum scale $Q$ increase
in strength as the center-of-mass energy increases, despite the
phenomenon of asymptotic freedom. The Rutherford form of the
parton-parton interaction is modified by a positive power of cms
energy, perhaps as large as 0.8. This could well lead to a novel regime
of strong-interaction phenomena in a kinematic regime naively expected
to be under control via QCD perturbation theory.

\begin{figure}[h!tb]
\leavevmode
\begin{center}
{\epsfxsize=3in\epsfysize=1.75in\epsfbox{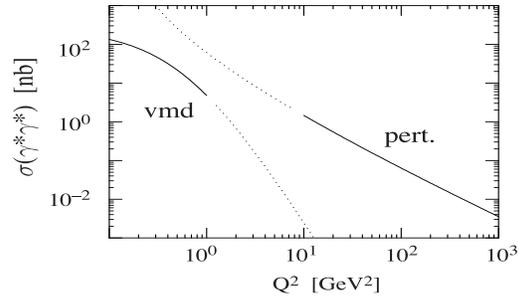}}
\end{center}
\caption{Estimated $\gamma^*-\gamma^*$ cross sections versus $Q^2$.}
\label{fig:e}
\end{figure}

The formalism underlying this involves the exchange of a ladder built
from gluons, the so-called ``hard Pomeron" or  ``BFKL
Pomeron"\cite{ref:D}.  A very clean prototype process is the
interaction of two spacelike virtual photons with each other at extreme
cms energy. The interaction of the two small color dipoles via one
gluon exchange is enhanced by extra gluon emission, and the expectation
is, in the limiting case of large $Q^2$ and very large $s$, that there
are no residual soft effects\cite{ref:E}.  This is in contrast to the
situation in $ep$ interactions, where the ``aligned-jet" mechanism
probably dominates the small-$x$ physics\cite{ref:F}.  In this case the
important fluctuations of the virtual photon into quark-antiquark are
those which create a large dipole-moment and small internal $p_t$ of
the $q\bar q$ system.  While the configuration is improbable at large
$Q^2$, the interaction with proton, when it occurs, is strong, leading
to scaling behavior and good phenomenological results. In collisions of
two virtual photons, however, the ``alignment" has to occur twice, once
for each photon, and therefore the cross-section falls as $Q^{-4}$,
while the perturbative piece scales as $Q^{-2}$. Numbers are being
provided by Brodsky, Hautmann and Soper\cite{ref:E};  beyond $10\
GeV^2$ the perturbative piece begins to dominate (Fig.~\ref{fig:e}).

Typical BFKL final states are multijet; in the $\gamma^*\gamma^*$ case
their $p_t$ are all supposed to be of order $Q$, and the mean rapidity
spacing of the gluon jets is inversely correlated (roughly) with the
exponent of the rise with energy of the cross section:
\begin{equation}
0.4\, \Delta\eta = \frac{12\alpha_s\Delta\eta}{\pi\, \ell n\, 2} \sim 1
\quad {\rm or}\quad \Delta\eta \sim 2.5 \ .
\end{equation}

In hadron-hadron collisions, the same phenomenon occurs, with gluons or
quarks, not virtual photons, as projectiles. As emphasized by Mueller
and Navelet\cite{ref:G},  there should be leading jets with $p_t$ of
order $Q$, the generic $p_t$ scale, to define the scale of the BFKL
process. And again there should be a multijet structure in between the
leading jets, with mean rapidity spacing again of order 2.5.

In order for the multijet phenomenon to occur (and presumably the BFKL
energy dependence as well!) very large $s/Q^2$ is therefore necessary. A
naive estimate of when the quark-quark cross section gets large (the
``black-quark threshold") is
\begin{equation}
\sigma\sim \frac{4\pi\, \alpha^2_s}{Q^2}\
\left(\frac{s}{Q^2}\right)^{0.4} \approx \frac{2\pi}{Q^2}
\end{equation}
or
\begin{equation}
\sqrt s \sim 100\ Q \ .
\end{equation}
The bottom line is that very large subenergies are preferred. 

\subsection{Rapidity Gaps}
\label{ssec:2c}

The presence of rapidity gaps in soft hadron-hadron collision processes
is an old story: this is the physics of elastic scattering and
diffraction dissociation. If, however, quarks can get ``black", then
one can anticipate a new class of diffractive processes that are
intrinsically short-distance. In addition there are hard diffractive
processes induced by color-singlet two-gluon exchange (or
electroweak-boson exchange) which also can lead to final states
containing both rapidity gaps and jets, even without invoking the BFKL
strong interaction.

The study of hard diffraction (rapidity gaps and jets in the same final
state) is in its infancy. It was proposed by Ingelman and
Schlein\cite{ref:H}, and dijets in single diffractive dissociation have
been found by Schlein and collaborators in the UA8
experiment\cite{ref:I}.  Since then rapidity gaps between coplanar
dijets have been observed in $p\bar p$ collisions by D0 and
CDF\cite{ref:J},  and gaps are seen as an important final state in deep
inelastic $ep$ collisions at HERA\cite{ref:K}.

So there are many such hard diffractive processes, some characterized
by a semisoft momentum transfer across the rapidity gap (UA8, HERA)
and others by a large momentum transfer (CDF, D0). Sorting it all out
will be a major program for the future. The soft and semisoft
diffraction is arguably a consequence of the blackness of the
constituent quark. It is a central question whether BFKL blackness of
partons is in fact just a smooth extrapolation to smaller scales of the
observed blackness of constituent quarks, or whether they are distinct
phenomena. The answer to this question will have to be data driven.

There is a lot which still needs to be done experimentally to study
well the various forms of hard (and for that matter semisoft and soft)
diffraction. Roman-pot detectors in both beam directions to catch
leading particles should be a standard supplement to every modern
barrel detector at colliders. At present there is only the beginning of
a realization of the value of this simple addition to the
instrumentation. In addition, for the optimal study of BFKL phenomena,
there should be good detection capability at large pseudorapidity (4 to
7 at the TeVatron, and 4 to 9 at the LHC) to identify the
BFKL-Mueller-Navelet tagging jets as well as to validate the presence
of rapidity gaps. The need for this large-acceptance capability is even
less appreciated;  I have advocated its importance for some
time\cite{ref:L} and can only report my total state of frustration.

In the shadowy world of diffractive phenomena, nuclei as targets or
projectiles are very valuable, since they are a way of tuning the
degree of blackness present in the collision process. The
$A$-dependence of the small-$x$ behavior in electroproduction, for
example, provides strong evidence for the aligned-jet picture of the
dynamics at fixed-target energies. $A$-dependence studies would be an
especially valuable tool at HERA energies, where there is clear
evidence for onset of new behavior, either BFKL or its precursor, in
the structure function $F_2$. The high observed gluon density in the
nucleon presages a much higher gluon density in nuclei. The recent
workshop studies at DESY\cite{ref:M} show that even for $Al$ and
certainly for $Pb$ the gluon density $xG(x)$ should be large enough to
exhibit saturation effects at attainable energy.

The study of rapidity gaps with ion projectiles at HERA would be an
especially useful way to sort out the various theoretical scenarios on
the diffractive mechanisms in electroproduction. For example most
models have anticipated that the typical diffracted mass would be of
order $Q$. The newest data clearly show a contribution from diffracted
masses large compared to $Q$. In the aligned-jet picture, the former
piece comes from elastic scattering of the slower quark or antiquark
from the nucleon ($A^{2/3}$), while the latter piece would be
diffraction dissociation ($A^{1/3}$). Other models will I am sure
differ from this expectation and, no matter what, the progress will
need to be data driven.

It seems to me that an $e$-ion capability at HERA is a most natural
upgrade path, one that flows naturally from the major contribution of
the present HERA program to strong-interaction physics, and one which
will surely be very productive and will help to complete the story
which HERA has so successfully initiated.

In any case the bottom line is that the $A$-dependence studies remain of
great value at collider energies: $e-A$ at HERA, and $p-A$ at RHIC and
LHC.

And the bottom lines for rapidity-gap physics are for me whether the
``new" BFKL physics is a smooth extension of the strong-coupling
physics of the constituent quarks, whether soft and hard diffraction
are smoothly connected, and what role (if any) the chiral limit of QCD
plays in these issues. It is possible that these issues may turn out to
be not distinct, but really just the same.

\subsection{The Chiral Phase in QCD; Disoriented Chiral Condensate}
\label{ssec:2d} 

The up and down quarks are at short distances almost massless. This
implies that QCD has a nearly exact $SU(2)\times SU(2)$ symmetry
corresponding to separate isospin rotations of left-handed and
right-handed quarks. This symmetry is spontaneously broken.  There is a
vacuum condensate (like the Higgs condensate) and the pions emerge as
collective (Goldstone-boson) modes of the condensate. In the perfect
symmetry limit the pions would be massless.

Given the mechanism of spontaneous symmetry breakdown, much of the
long-distance, low-energy limit of the theory is rather well
determined, in particular the low-energy limit of the interactions of
the pions with matter and with each other. This is codified in the
``chiral effective Lagrangian," with the degrees of freedom being
constituent quarks and pions. The validity of this effective theory
extends from zero energy up to a mass scale of 500-1000
$MeV$\cite{ref:N}.

It is a challenge for QCD theorists to derive the existence of this
chiral phase from first principles. There is evidence from the lattice
calculations that a chiral condensate forms, but the mechanism remains
unclear. There is also an interesting line of work by Shuryak,
Diakonov, and others which argues that the chiral symmetry breaking
mechanism can be traced to the presence of instantons in the QCD
vacuum\cite{ref:O}.  This idea is discussed more in my SLAC Summer
Institute talk\cite{ref:A}.

But while everyone (theorists) talks about the QCD vacuum, hardly
anyone (experimentalists) tries to do something about it. For the last
three years some of us have banded together to do a test/experiment at
the Tevatron collider (T864 (MiniMax)) to search for something called
disoriented chiral condensate (DCC)\cite{ref:P}.

What is DCC? It is a conjectured piece of strong-interaction vacuum
with an unusual orientation of its chiral order parameter. The vacuum
condensate associated with the $SU(2) \times SU(2) = 0(4)$
chiral-symmetry spontaneous breaking is a chiral four-vector ($\sigma,
\vec\pi$) which in normal vacuum points in the sigma direction. But
inside a hot fireball shell created in a high-energy collision, the
chiral orientation need not be the same. If it is different, \eg\
points in the $\pi^0$ direction, then this piece of wrongly oriented
vacuum will eventually decay into true vacuum with emission of a
semiclassical pulse of $\pi^0$'s. In other events the chiral order
parameter might point in a charged-$\pi$ direction and only charged
pions will be found in the final state. The experimental signature is
large event-to-event fluctuations of the fraction of produced pions
which are neutral. One finds
\begin{equation}
f = \frac{N_{\pi^0}}{N_{\pi^0}+N_{\pi^+}+N_{\pi^-}}
\end{equation}
\begin{equation}
\left(\frac{dN}{df}\right)_{\rm DCC} = \frac{1}{2\sqrt f}
\end{equation}
\begin{equation}
\left(\frac{dN}{df}\right)_{\rm Generic} \approx \delta
\left( f-\frac{1}{3}\right) \ .
\end{equation}
There are other possible signatures as well. DCC which is produced at
large transverse velocity may be easiest to find. Or cutting on low
transverse momentum or groups of pions with low relative transverse
momentum are other possibilities.

This theoretical picture is also motivated by cosmic-ray events
(Centauro, anti-Centauro) seen in mountaintop and balloon-borne
emulsion chambers\cite{ref:Q}.  In the Chacaltaya events, it is claimed
that groups of hadrons which exhibit the Centauro-like behavior also
have low relative transverse momentum, perhaps in line with the above
picture, which suggests that the hadrons are emitted at late proper
times from a large emitting area.

Our test/experiment T864 was proposed in April 1993, and is now
completed. This is not the occasion to describe it in any detail.
Suffice it to say that we have recorded about 8 million events, and
have initiated the data  analysis. We have found a promising analysis
technique\cite{ref:R}  which utilizes ratios of bivariate factorial
moments (standard tools of the trade in the multiparticle-dynamics
community) to finesse many (not all) of the serious efficiency problems
faced in such a search---especially ours, which uses unsophisticated
apparatus of small acceptance.

It is still very early in the analysis. Thus far we see no evidence for
spectacular events \textit{a la} Centauro and JACEE, although we need
to do more work to assess the level of significance. The
factorial-moment method shows consistency with generic pion production,
with a DCC admixture limited to something like 10 to 20 percent of the
generic production (although it is too early to really quote numbers).

No matter what comes out, we have learned a lot about how to go
about searching for DCC, and believe that the search can and should be
done with better detectors and improved analysis technique. We stand
ready to help others make the search. At present there is a growing
interest within the nuclear-physics, heavy-ion community in searching for
DCC. I hope this might happen in the high-energy community as well.

\subsection{Underlying-event and Minimum-bias Physics}
\label{ssec:2e}

The physics of mundane, minimum-bias events, and the underlying-event
portion of high-$p_t$ multijet events is definitely not a glamour
subject. Nevertheless it is a topic which is important in its own right
as well as having serious engineering value in the interpretation of
the high-$p_t$, high glamour physics.

The data base in electron-positron collisions is by now quite complete,
and the theoretical descriptions relatively mature. But even for this
case there are new challenges appearing, especially in the analysis of
$WW \rightarrow q\bar q\,q\bar q$ final states at LEP II. One cannot
superpose the final-state hadron distributions from two independent $W$
decays, because the hadronization of each jet pair occurs
simultaneously in overlapping regions of space.  Many of the
final-state-interaction properties are nonperturbative in origin and
will be a challenge to QCD phenomenology. And the stakes are high;
namely, accurate measurement of the $W$ mass\cite{ref:S}.

But the situation is the worst in hadron-hadron collisions\cite{ref:T}.
The main minimum-bias data base at collider energies is limited to
UA(1), where a small band of analysts carry on valiantly an analysis of
that old data, and UA(5)---a nonmagnetic streamer-chamber experiment
with low gamma-ray efficiency---which nevertheless to this day remains
one of the most serious sources of real information.  There is a small
amount of data from our much-too-modest predecessor experiment at C0,
E(735). CDF and D0 have poor capability at transverse momenta under 400
$MeV$, greatly hampering meaningful minimum-bias analyses. In addition,
D0 of course does not have magnetic analysis, also a serious
limitation. Much of the minimum-bias physics at collider energies is
reduced to Monte-Carlo cocktails, the quality of which, for an outsider
like me, is hard to digest. Of course, the creators do a great service
to the field. Nevertheless what is really assumed? What is the real
data base that is used, and what are the limits of applicability? Were
a real hadron-hadron minimum-bias data base to suddenly appear, how
well would those codes really do?

With the great investment being made in learning about the rare
processes, it would seem especially prudent to make at least some
modest investment toward understanding the more common processes. To do
that job well requires a community of interest able to mount a
dedicated effort with specialized detectors that have low-$p_t$
sensitivity, acceptance large enough to observe the final-state energy
(not $E_T$) distribution, along with some particle identification. Such
detectors should acquire a large database commensurate with modern
data-acquisition capabilities.  It is with despair that I note that
there is insufficient interest within this country for this to happen.
It is not only a pity that this is the case, it is bad science.

\section{Matching the Physics to the Facilities}
\label{sec:3}

\subsection{Electron-Positron Colliders}
\label{ssec:3a}

As we have already mentioned, the decay of virtual $\gamma$ and $Z$
into jets plus gluons is well studied and can be extrapolated safely to
higher energies. The main QCD phenomenon for higher energies that is not
presently under study is the BFKL hard Pomeron. Ideally one wants to
study the final-state hadron system in the high-energy collision of two
virtual photons. For this one must tag the secondary electron and
positron, which is not at all easy. But in addition one would like to
see the leading hadrons and Mueller-Navelet tagging jets. While finding
the electrons may be doable if care is exercised in the initial design
of the final focus system, having low $\beta^*$ and seeing the
leading-particle hadrons looks very hard to me. Higher $\beta^*$ (if
the luminosity loss is tolerable) and/or a separate detector/collision
region may be necessary.

\subsection{Electron-Proton Colliders (HERA, eventually LHC?)}
\label{ssec:3b}  

Options for the future of HERA are now under study. While this is an
issue outside the scope of the Snowmass charge, future US participation
will be influenced by the issues and the unique physics opportunities.
At present the HERA program probes in powerful and unique ways the
mechanisms of diffraction and the nature of the Pomeron.

\begin{figure}[h!tb]
\leavevmode
\begin{center}
{\epsfxsize=3.15in\epsfbox{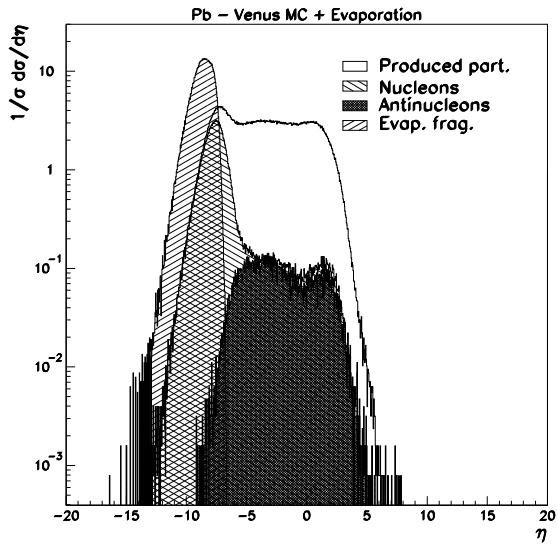}}
\end{center}
\caption{Hadron production spectra for $e-A$ collisions at HERA.}
\label{fig:g}
\end{figure}

One of the options for HERA is to study electron-ion collisions. To me
this is a very attractive option, which would, as mentioned already in
the previous section, consolidate the gains already made in
understanding the nature of diffractive processes and the physics
behind the rise of the deep-inelastic structure function at very small
$x$. Together with such a program it would be very natural to increase
the acceptance of the present detectors in the proton direction to more
fully interpret the diffractive phenomena. This is a deficiency already
apparent in the electron-proton collisions now being studied. Forward
detectors, within and outside the beam-pipe acceptance, are needed to
define well the rapidity gaps. The proton fragmentation, \eg\ into
tagging jets, is an important signature for BFKL dynamics as well. The
deficiency in the present coverage (Fig.\ref{fig:g}) is serious,
especially if one is to study at all the nuclear fragmentation.

\subsection{Hadron-Hadron Colliders}
\label{ssec:3c}  

We have already mentioned two basic QCD frontiers in hadron-hadron
collisions which impact in a most fundamental way on detector design.
One is the high-granularity frontier of very small lego-pixel
resolution to look in the interior of jets. Here by necessity the
burden in reconstruction of jets-within-jets goes onto the
electromagnetic calorimetry and the charged-particle tracking. Hadron
calorimetry is almost certainly too coarse to be definitive in
resolving the substructure. But I believe multijet spectroscopy can
still be done without the hadron calorimetry. The price paid is some
inefficiency per jet of order 20-30\%, which occurs if it is lost into
a lower $p_t$ bin when too much transverse energy goes into $K_L$ or
neutrons. This strategy would of course need careful study, and even if
it works in principle there is a great demand on the detector design.
But, assuming the strategy is found in principle to be okay, the
physical limit on resolving individual $\gamma$'s or tracks would be at
the few millimeter level, leading in typical detectors the possibility
of exploring extended phase space to rapidities of 3 to 4. Providing
this texture in all of phase space would involve an enormous number of
readout channels. But it seems to me it would be valuable to instrument
at least a portion of the large-angle acceptance in such a way.

Another frontier that just has to be useful is that of particle
identification. There has not been a Cerenkov detector in a
hadron-hadron collider since the ISR. The emergence of collider
detectors to do heavy flavor physics should change that, and there may
be many other useful QCD byproducts coming from that extended
capability.

Yet another frontier is simply to supply the capability for studying
low-$p_t$ phenomena well; here low-$p_t$ means down to tens of $MeV$
per particle. Novel phenomena like DCC production or other particle
production mechanisms which occur well within the light cone (instead
of very near it) may leave their signature in the properties of
low-$p_t$ secondaries. So far CDF and D0 have exhibited neither the
capability nor much interest in exploration in this direction.

Finally there is the issue of the extension of acceptance into the
forward region. Magnetic analysis of charged particles should extend
beyond the barrel-solenoid region all the way to the leading particle
region. The problem of peaceful coexistence with the beam-pipe
showering is a serious one, but one that I believe can be solved. And,
as already mentioned, Roman-pot detectors within the beam-pipe
acceptance should be designed {\it ab initio} into the machine lattice,
not only to see the elastic and diffractive protons, but also to pick
off inelastic leading hadrons which cannot be found with detectors
exterior to the pipe. Again, to do all this requires an attitude by an
interested community that this is indeed not only worth doing, but is a
necessary adjunct to the highest priority, high-$p_t$ program of
exploratory physics.

\subsection{Possible New Options: (CO at the TeVatron, and FELIX at
LHC)} 
\label{ssec:3d}

There are at present two fresh options for innovations. One is at
Fermilab, where there are preliminary studies initiated by its director
to explore the physics case for---and feasibility of---upgrading the C0
collision region for a major third detector. The physics focus would be
on high-yield charm and bottom physics, the latter at or beyond what is
needed to observe CP-violating phenomena. This is a right and proper
central goal. However there is the possibility that this lead program
could be supplemented with full-acceptance capability for studying some
of the topics mentioned above.

\begin{figure}[htb!]
\leavevmode
\begin{center}
{\epsfxsize=3in \epsfbox{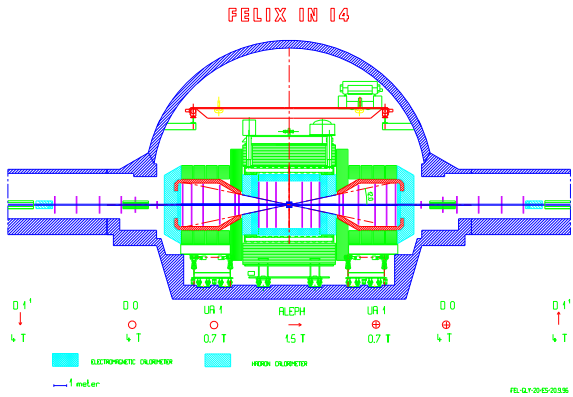}}
\end{center}
\caption{Layout of the FELIX detector for the LHC.}
\label{fig:h}
\end{figure}

A study is underway under the leadership of Jeff Appel and Peter
Garbincius. The time scale is very short, because there is a window of
opportunity for doing the civil construction during a
Main-Injector-commissioning shutdown. Up-to-date information is best
obtained by consulting Jeff and/or Peter. Additional insight may be
gleaned from the public report of Fermilab's PAC\cite{ref:U}, which
encourages nothing at C0 except heavy flavor physics.  In my opinion,
this attitude is deplorable.

The FELIX initiative at the LHC aims to provide a true full-acceptance
detector which would be the definitive QCD facility for that program.
It would be located at intersection region I4, where ALEPH now resides,
and in fact would use its solenoid as its central magnet. The next
magnetic stages upstream and downstream would utilize the UA1 magnet
yokes (with a new coil), which are modular and ``portable"
(Fig.~\ref{fig:h}). Thus the central free space of $\pm$ 10 meters
would be well covered by magnetic field appropriate for full magnetic
analysis of charged secondaries with $\eta \leq 6$. Magnetic analysis
in the region from 10 to 100 meters from the collision point would be
provided by machine magnets and appropriate tracking, sufficient to
provide complete rapidity coverage for the charged particles of $\eta
\geq 6$, including the elastic and diffractive protons.

Because there is little cost in civil construction and
magnetic/calorimetric tonnage, the main cost for the detector is
measured in terms of number of readout channels. A base cost is what is
needed to bring the beams together into collision. With a multiplier of
two to three of this base cost, a great deal of Stage I physics could
be accomplished.

More details on this initiative can be obtained from Karsten Eggert
(CERN) or Cyrus Taylor (Case Western Reserve University), and by
consulting the FELIX web page http://www.cern.ch/FELIX.

%

\end{document}